\documentclass[final]{aipproc}

\layoutstyle{6x9}
\SetInternalRegister\hbadness{8000} % pseudo latin isn't breaking very well :-)

\def\be{\begin{equation}}
\def\ee{\end{equation}}
\def\bea{\begin{eqnarray}}
\def\eea{\end{eqnarray}} 
\def\ba{\begin{eqnarray}}
\def\ea{\end{eqnarray}} 

\begin{document}

\title{A quark loop model for heavy mesons\footnote{Invited talk 
at QCD@Work, Martina Franca, Italy, June 16-20 2001.}}


\author{Aldo Deandrea}{
  address={IPN, Universit\'e de Lyon I, 4 rue Enrico~Fermi, 
F-69622 Villeurbanne Cedex, France},
  email={deandrea@ipnl.in2p3.fr},
  thanks={}
}

\copyrightholder{}
\copyrightyear{2001}

\begin{abstract}
I consider a model based on a
quark--meson interaction Lagrangian. The transition amplitudes
are evaluated by computing diagrams in which heavy and light mesons
are attached to quark loops. The light chiral symmetry 
relations and the heavy quark spin-flavour symmetry dictated by the
heavy quark effective theory are implemented. 
The model allows to compute the decay form factors and therefore can give
predictions for the decay rates, the invariant mass spectra and the
asymmetries. 
\end{abstract}

\date{\today}

\maketitle

\section{Introduction}
The increasing number of available data on heavy meson processes demands
theoretical predictions for these processes to be compared with experiment. 
I consider a simple model, based on an effective constituent quark-meson
Lagrangian containing both light and heavy degrees of freedom, constrained
by the known symmetries of QCD in the limit $m_Q\to\infty$ and the 
light chiral symmetry relations. I write a Lagrangian at the
meson-quark level \cite{nc}. This allows
to deduce from a small number of parameters the heavy meson couplings and form
factors, with a considerable reduction in the number of free parameters with
respect to the Lagrangian written in terms of meson fields only
\cite{Casalbuoni:1997pg}.

The part of the quark-meson effective Lagrangian involving heavy and 
light quarks and heavy mesons is:
\begin{eqnarray}
{\cal L}_{h \ell}&=&{\bar Q}_v i v\cdot \partial Q_v
-\left( {\bar \chi}({\bar H}+{\bar S}+ i{\bar T}_\mu
{D^\mu \over {\Lambda_\chi}})Q_v +h.c.\right)\nonumber \\
&+&\frac{1}{2 G_3} {\mathrm {Tr}}[({\bar H}+{\bar S})(H-S)]
+\frac{1}{2 G_4}
{\mathrm {Tr}} [{\bar T}_\mu T^\mu ] \label{qh1}
\end{eqnarray}
where $Q_v$ is the effective heavy quark field, $\chi$ is the light quark 
field, $G_3$, $G_4$ are coupling constants
and $\Lambda_\chi$ ($= 1$ GeV) is a dimensional parameter. 
The Lagrangian (\ref{qh1}) is heavy spin and flavour symmetric. 
Note that the fields $H$ and $S$ have the same coupling constant. 
By putting these two coupling constants equal, one assumes that the effective
quark-meson Lagrangian can be obtained from a four quark interaction of
the NJL type \cite{qmi}.

The cut-off prescription is part of the dynamical 
information regarding QCD which is introduced in the model.
The idea is to mimic the QCD behaviour in a simple and calculable way. 
In the infrared the model is not confining and 
its range of validity can not be extended below energies of the order of
$\Lambda_{QCD}$. In practice one introduces an infrared cut-off $\mu$, 
to take this into account.

Models related to the one discussed here, with different regularization
prescriptions and different approaches are \cite{bardeen,holdom}. 
The cut-off prescription used here is implemented via a proper time
regularization. After continuation to the Euclidean it reads, for the
light quark propagator: 
\begin{equation}
\int d^4 k_E \frac{1}{k_E^2+m^2} \to \int d^4 k_E \int^{1/
\Lambda^2}_{1/\mu^2} ds\; e^{-s (k_E^2+m^2)}\label{cutoff}
\end{equation}
where $\mu$ and $\Lambda$ are infrared and ultraviolet cut-offs.

The cut-off prescription is similar to the one used in \cite{qmi}, with
$\Lambda=1.25$ GeV; the numerical results are not strongly dependent on the
value of $\Lambda$. The constituent mass $m$ in the NJL models represents the
order parameter discriminating between the phases of broken and unbroken chiral
symmetry and can be fixed by solving a gap equation, which gives $m$ as a
function of the scale mass $\mu$ for given values of the other parameters. 
Here I take $m=300$ MeV and $\mu=300$ MeV.

\section{Heavy-to-Heavy Form Factors}

As an example of the quantities that can be analytically calculated
in the model, one can examine the Isgur-Wise function $\xi$:
\begin{equation}
\langle D(v^\prime)|{\bar c} \gamma_\mu (1-\gamma_5) b|
B(v)\rangle = \sqrt{M_B M_D} C_{cb}\; \xi(\omega) (v_\mu + v^{\prime }_{\mu})
\end{equation}
where $\omega= v \cdot v^\prime$ and $C_{cb}$ contains
logarithmic corrections depending on $\alpha_s$; within the approximations
used here, it can be put equal to 1. At leading order $\xi(1)=1$.
The same universal function $\xi$ also parameterizes $B \to D^*$ semileptonic
decay. One finds:
\begin{equation}
\xi(\omega)=Z_H \left[ \frac{2}{1+\omega} I_3(\Delta_H)+\left( m+\frac{2
\Delta_H}{1+\omega} \right)
I_5(\Delta_H, \Delta_H,\omega) \right] ~~.\label{xi}
\end{equation}
where:
\begin{eqnarray}
I_3(\Delta) &=& - \frac{iN_c}{16\pi^4} \int^{\mathrm {reg}}
\frac{d^4k}{(k^2-m^2)(v\cdot k + \Delta + i\epsilon)}\nonumber \\
&=&{N_c \over {16\,{{\pi }^{{3/2}}}}}
\int_{1/{{\Lambda}^2}}^{1/{{\mu }^2}} {ds \over {s^{3/2}}}
\; e^{- s( {m^2} - {{\Delta }^2} ) }\;
\left( 1 + {\mathrm {erf}} (\Delta\sqrt{s}) \right)\\
I_5(\Delta_1,\Delta_2,\omega) & = & \frac{iN_c}{16\pi^4} \int^{\mathrm {reg}}
\frac{d^4k}{(k^2-m^2)(v\cdot k + \Delta_1 + i\epsilon )
(v'\cdot k + \Delta_2 + i\epsilon )} \nonumber \\
 & = & \int_{0}^{1} dx \frac{1}{1+2x^2 (1-\omega)+2x
(\omega-1)}\times\nonumber\\
&&\Big[ \frac{6}{16\pi^{3/2}}\int_{1/\Lambda^2}^{1/\mu^2} ds~\sigma
\; e^{-s(m^2-\sigma^2)} \; s^{-1/2}\; (1+ {\mathrm {erf}}
(\sigma\sqrt{s})) +\nonumber\\
&&\frac{6}{16\pi^2}\int_{1/\Lambda^2}^{1/\mu^2}
ds \; e^{-s(m^2-2\sigma^2)}\; s^{-1}\Big]
\end{eqnarray}
In these equations
\begin{equation}
\Gamma(\alpha,x_0,x_1) = \int_{x_0}^{x_1} dt\;  e^{-t}\; t^{\alpha-1}
\end{equation}
is the generalized incomplete gamma function, erf is the error function and
\begin{equation}
\sigma(x,\Delta_1,\Delta_2,\omega)={{{\Delta_1}\,\left( 1 - x \right)  +
{\Delta_2}\,x}\over {{\sqrt{1 + 2\,\left(\omega -1  \right) \,x +
2\,\left(1-\omega\right) \,{x^2}}}}}~.
\end{equation}
One can compute in a similar way the form factors describing 
the semi-leptonic decays of a meson belonging to the fundamental negative
parity multiplet $H$ into the positive parity mesons in the $S$ and $T$ 
multiplets \cite{nc}. Examples of these decays are $B \rightarrow D^{**} l \nu$
where $D^{**}$ can be either a $S$ state or a $T$ state.
These decays are described by two form
factors $\tau_{1/2}, \tau_{3/2}$ \cite{IW2} which
can be computed in the model by a loop calculation similar to the one used to
obtain $\xi(\omega)$ \cite{nc,Deandrea:1999pa}. 

The numerical results for the form factors are in Table \ref{tab:3tab}. 
\begin{table}
\caption{Form factors and slopes. $\Delta_H$ in GeV.}
\label{tab:3tab}
\vspace{0.2cm}
\begin{tabular}{ccccccc} 
\hline
{$\Delta_H$} & {$\xi(1)$} & 
{$\rho^2_{IW}$} &{$\tau_{1/2}(1)$} &
{$\rho^2_{1/2}$} &{$\tau_{3/2}(1)$} &
{$\rho^2_{3/2}$} \\
\hline
0.3 & 1 & 0.72 & 0.08 & 0.8 & 0.48 & 1.4 \\
0.4 & 1 & 0.87 & 0.09 & 1.1 & 0.56 & 2.3 \\
0.5 & 1 & 1.14 & 0.09 & 2.7 & 0.67 & 3.0 \\
\hline
\end{tabular}
\end{table}
\vspace*{3pt}
The predictions for a few branching ratios calculated in the model are given in
Table \ref{tab:sl}.
\begin{table} [htb]
\label{tab:sl}
\hfil
\vbox{\offinterlineskip
\halign{&#& \strut\quad#\hfil\quad\cr
\hline
&Decay mode&& $\Delta_H=0.3$&& $\Delta_H=0.4$&& $\Delta_H=0.5$&&
Exp. \cite{PDG}& \cr
\hline
&${B^0}\to D\ell\nu $&& $3.0$&&$2.7$&&$2.2$&&$2.10 \pm 0.19$& \cr
&${B^0}\to D^{*}\ell\nu$&& $7.6$&& $6.9$&& $5.9$&& $4.60\pm 0.27$ &\cr
&${B^0}\to D_0\ell\nu$&& $0.03$&& $0.005$&& $0.003$&& -- &\cr
&${B^0}\to D^{*\prime}_{1}\ell\nu$&& $0.03$&& $0.008$&& $0.0045$&&-- &\cr
&${B^0}\to D_{1}^*\ell\nu$&& $0.27$&& $0.18$&& $0.13$&& $<0.74$ &\cr
&${B^0}\to D^{*}_{2}\ell\nu$&& $0.43$&& $0.34$&& $0.30$&& $<0.65$&\cr
\hline}}
\caption{Branching ratios (\%) for semileptonic $B$ decays.
Theoretical predictions for three
values of $\Delta_H$ and experimental results.
Units of $\Delta_H$ in GeV.}
\end{table}

\section{Heavy-to-Light Form Factors}
The model allows to compute the $B$ semileptonic decay form factor to $\pi$,
$\rho$, etc. The form factors of $B$ to a vector meson $V$ consist of two kind
of contributions. In the first one the current is
directly attached to the loop of quarks. In the second, there is a
intermediate state between the current and the $B\; V$ system 
\cite{Deandrea:1999ww}. For $B \to \pi$ form factors an extra
contribution is also taken into account \cite{Deandrea:2000wq}.
Results are in good agreement with available data.  
For $B\to \pi \ell\nu$ (using $V_{ub}=0.0032$, 
$\tau_B=1.56 10^{-12}$ s):
\begin{equation}
{\cal B}(\bar B^0 \to \pi^+ \ell \nu) = (1.1\pm 0.5) \times 10^{-4} \; ,
\end{equation}
for $B\to \rho \ell\nu$:
\begin{equation}
{\cal B}(\bar B^0 \to \rho^+ \ell \nu) = (2.5\pm 0.8) \times 10^{-4} \; ,
\end{equation}
for $B\to a_1 \ell\nu$:
\begin{equation}
{\cal B}(\bar B^0 \to a_1^+ \ell \nu) = (8.4 \pm 1.6) \times 10^{-4} \; . 
\end{equation}

In the limit of heavy mass for the initial meson and of large energy 
for the final one (LEET), the expressions of the form factors simplify
and for $B \to V l \nu$, they reduce only to two
independent functions \cite{leet}. The
four-momentum of the heavy meson is written as $p= M_H v$ in terms of the
mass and the velocity of the heavy meson. The four-momentum of the light
vector meson is written as $p'=E n$ where $E=v \cdot p'$ is the energy of the
light meson and $n$ is a four-vector defined by $v \cdot n=1, n^2=0$. 
The relation between $q^2$ and $E$ is:
\begin{equation}
q^2=M_H^2- 2 M_H E +m_V^2
\end{equation}
The large energy limit is defined as : 
\be
\Lambda_{QCD}, m_V << M_H, E
\ee
keeping $v$ and $n$ fixed and $m_V$ is the mass of the light vector meson. 
The relations between the form factors appearing in the LEET limit 
constitute a powerful theoretical cross--check of the formulas derived
in the model. The result is as follows:  
\ba
A_0(q^2)&=&\left(1-\frac{m_V^2}{M_H E}\right)\zeta_{||}(M_H,E)
+\frac{m_V}{M_H}\,\zeta_{\perp}(M_H,E) \label{fatta0}\\
A_1(q^2)&=&\frac{2E}{M_H + m_V}\,\zeta_{\perp}(M_H,E) \label{fatta}\\
A_2(q^2)&=&\left(1+\frac{m_V}{M_H}\right)\left[\zeta_{\perp}(M_H,E)-
\frac{m_V}{E}\zeta_{||}(M_H,E)\right]\\
V(q^2)&=&\left(1+\frac{m_V}{M_H}\right)\zeta_{\perp}(M_H,E).
\label{fattv}
\ea
The explicit expressions for $\zeta_{||}$ and $\zeta_{\perp}$
are \cite{moriond}:
\ba
\zeta_{||}(M_H,E)&=&\frac{\sqrt{M_H Z_H}\; m_V^2}{2 E f_V} \Big[
I_3\left(\frac{m_V}{2}\right)-I_3\left(-\frac{m_V}{2}\right)
\nonumber \\
&+&4 \Delta_H m_V\; Z
\Big] \sim \frac{\sqrt{M_H}}{E} 
\label{zetapar}
\ea
\be
\zeta_{\perp}(M_H,E)=\frac{\sqrt{M_H Z_H}\; m_V^2} {2 E f_V}
\left[I_3(\Delta_H) + m_V^2 \; Z \right]
\sim \frac{\sqrt{M_H}}{E}\; ,
\label{zetaperp}
\ee
where terms proportional to the constituent light quark mass $m$ have been
neglected. It is interesting to note that in LEET one can also
relate the tensor form factor $T_1$, $T_2$ and $T_3$ to the semileptonic ones
and to the $\zeta_{\perp}$ and $\zeta_{||}$ form factors of the LEET limit
\cite{leet}: 
\begin{eqnarray} 
T_1(q^2)&=&\zeta_{\perp}(M_H,E)\,,\\
T_2(q^2)&=&\left(1-\frac{q^2}{M_H^2-m_V^2}\right)\zeta_{\perp}(M_H,E)\,,\\
T_3(q^2)&=&\zeta_{\perp}(M_H,E)-\frac{m_V}{E}\left(1-\frac{m_V^2}{M_H^2}\right)
\zeta_{||}(M_H,E)\,.
\label{fattt3}
\end{eqnarray}
$\zeta_{\perp}$ and $\zeta_{||}$ obtained in this way agree with those of
(\ref{zetapar},\ref{zetaperp}) \cite{Deandrea:2001qs}.
Concerning the scaling properties of $\zeta_{||}$ and 
$\zeta_{\perp}$, the asymptotic $E$-dependence is not predicted by the
large energy limit. As $E \sim M$ at $q^2=0$ the Feynman mechanism 
contribution to the form factors would indicate a $1/E^2$
behaviour rather than the $1/E$ found in the model. Note however that
the $E$-dependence is not rigorously established in QCD.

\section{Conclusions}
Calculating directly from the QCD
Lagrangian remains an extremely difficult task, in spite of the impressive
success of lattice QCD calculations. A most promising approach is the one based
on heavy meson effective Lagrangians, which incorporate the heavy quark
symmetries and in addition the approximate chiral symmetry for light quarks.
Although with increasing data such an approach is the best one beyond
direct QCD calculations, a large number of parameters have to be fixed
before obtaining predictions. An intermediate approach consists in using the
effective Lagrangian at the level of mesons and constituent quarks plus few
simple assumptions on the QCD dynamics. It allows to compute meson transition
amplitudes by evaluating loops of heavy and light quarks. The model
describes a number of essential features of heavy meson physics in a simple and
compact way, in particular Isgur-Wise scaling in the heavy-to-heavy
semileptonic decays and the large energy limit for the heavy-to-light ones.

\section*{Acknowledgments}
I would like to thank R. Gatto, G. Nardulli and  A.D. Polosa for the
fruitful scientific collaboration on which this work is based.  
Particular thanks go to the organizers of this workshop for inviting me
and for the pleasant atmosphere in Martina Franca.
Institut de Physique Nucl\'eaire de Lyon (IPN Lyon) is UMR 5822.

\end{document}